\documentclass[%
 reprint,%
 amssymb, amsmath,%
 aip,cha,%linenumbers%
]{revtex4-1}
\usepackage[columnwise]{lineno}

\usepackage{docs}%
\usepackage{bm}%
\usepackage[colorlinks=true,linkcolor=blue]{hyperref}%
\expandafter\ifx\csname package@font\endcsname\relax\else
 \expandafter\expandafter
 \expandafter\usepackage
 \expandafter\expandafter
 \expandafter{\csname package@font\endcsname}%
\fi
\hypersetup{
	colorlinks=true,
	linkcolor=blue,
	filecolor=magenta,}
\usepackage{graphicx}
\usepackage{lipsum}
\begin{document}
%\linenumbers

\title{Enhanced anisotropy and study of magnetization reversal in Co/C$_{60}$ bilayer thin film} %

\author{Srijani Mallik}%
\affiliation{Laboratory for Nanomagnetism and Magnetic Materials (LNMM), School of Physical Sciences, National Institute of Science Education and Research (NISER), HBNI, P.O.- Jatni, 752050, India}%
\author{Purbasha Sharangi}%
\affiliation{Laboratory for Nanomagnetism and Magnetic Materials (LNMM), School of Physical Sciences, National Institute of Science Education and Research (NISER), HBNI, P.O.- Jatni, 752050, India}%
\author{Biswajit Sahoo}%
\affiliation{Laboratory for Nanomagnetism and Magnetic Materials (LNMM), School of Physical Sciences, National Institute of Science Education and Research (NISER), HBNI, P.O.- Jatni, 752050, India}%
\author{Stefan Mattauch}%
\affiliation{J\"{u}lich Centre for Neutron Science (JCNS), Heinz Maier-Leibnitz Zentrum (MLZ), Forschungszentrum J\"{u}lich GmbH, Lichtenbergstr. 1, 85748 Garching, Germany}%
\author{Thomas Br\"{u}ckel}%
\affiliation{J\"{u}lich Centre for Neutron Science (JCNS), Heinz Maier-Leibnitz Zentrum (MLZ), Forschungszentrum J\"{u}lich GmbH, Lichtenbergstr. 1, 85748 Garching, Germany}%
\affiliation{PGI-4: Scattering Methods Forschungszentrum Julich GmbH 52425 Jülich, Germany}%
\author{Subhankar Bedanta}%
\email{sbedanta@niser.ac.in}
\affiliation{Laboratory for Nanomagnetism and Magnetic Materials (LNMM), School of Physical Sciences, National Institute of Science Education and Research (NISER), HBNI, P.O.- Jatni, 752050, India}%
%\date{March 2019}%
%\revised{November 2019}%

%\tableofcontents

\begin{abstract}
The interface between organic semiconductor [OSC]/ferromagnetic [FM] material can exhibit ferromagnetism due to their orbital hybridization. Charge/spin transfer may occur from FM to OSC layer leading to the formation of `spinterface' i.e. the interface exhibiting a finite magnetic moment. In this work, the magnetic properties of Co/C$_{60}$ bilayer thin film have been studied to probe the interface between Co and C$_{60}$ layer. Polarized neutron reflectivity [PNR] measurement indicates that the thickness and moment of the spinterface are $\sim$ 2 $\pm$ 0.18 nm and 0.8 $\pm$ 0.2 $\mu_B$/cage, respectively.
The comparison of the magnetization reversal between the Co/C$_{60}$ bilayer and the parent single layer Co thin film reveals that spinterface modifies the domain microstructure. Further, the anisotropy of the bilayer system shows a significant enhancement ($\sim$ two times) in comparison to its single layer counterpart which is probably due to an additional interfacial anisotropy arising from the orbital hybridization at the Co/C$_{60}$ interface.
\end{abstract}

\maketitle

Study of spin-dependent interfacial properties of ferromagnetic [FM]/organic semiconductor [OSC] interfaces have received immense research interest in the last decade \cite{Dediu2009,Atodiresei2010,Barraud2010}. This is due to both their potential applications in organic spintronic devices as well as the fundamental properties observed upon forming the `spinterface'. Spinterface is the gateway between two layers through which polarized spins can be transferred. It is a special type of interface having properties which are significantly different from their constituent layers \cite{Sanvito2010,Djeghloul2013,Steil}. The coupling between a FM and OSC molecule at the interface may promote FM state in the OSC layer \cite{Moorsom2014,Tran2011,Tran2013,Mallik2018,Djeghloul2016}. Buckminsterfullerene (C$_{60}$) is a potential candidate for such spin polarized charge transfer as it is composed of only carbon atoms. This is because, it exhibits properties like low spin-orbit coupling (due to lower atomic weight) and correspondingly longer spin lifetime, stability under thermal and mechanical duress, a reasonable mobility of 11 cm$^2$/V s etc. \cite{Moorsom2014,Tran2011,Tran2013,Mallik2018,Djeghloul2016}. The C atoms present in C$_{60}$ may undergo \textit{s}-\textit{p} hybridization mechanism to lower the total energy of the system. Thus, it is prone to hybridize orbitals with FM materials having unfilled \textit{d} orbitals. For example, Moorsom \textit{et al}., have studied Co/C$_{60}$ multilayers by polarized neutron reflectometry and found that magnetic moment can be induced in C$_{60}$ due to charge transfer \cite{Moorsom2014}. However, the exact thickness of the spinterface has not been deduced in their study because of interdiffusion between Co and C$_{60}$. Similarly, C$_{60}$ monolayers on Fe (001) reveal magnetic polarization of C$_{60}$ due to hybridization of C$_{60}$ and Fe orbitals \cite{Tran2011}. The hybrid interface between Fe and C$_{60}$ leads to magnetic moments $\mu_S$ = −0.21 and −0.27 $\mu_B$ per C molecule for C$_{60}$ layers on Fe(001) and Fe/W(001), respectively \cite{Tran2013}. Recently, spinterface of about 2 nm thickness in Fe/C$_{60}$ bilayer system has been observed which exhibits moment $\sim$ 1.5 to 3 $\mu_B$ per cage of C$_{60}$ \cite{Mallik2018}. The reason behind the charge transfer between the FM and the OSC layer is the $d_{FM}$ - $p_{OSC}$ orbital hybridization at the interface. It has been also reported that such spinterfaces have significant effect on the magnetization reversal mechanism and domains of epitaxial Fe/C$_{60}$ bilayer systems \cite{Mallik2018}. The aforementioned effects have not been explored so far for Co/C$_{60}$ systems.
\\ It should be noted that the hybridization between the FM and OSC molecules directly involve the orbitals, therefore, it is expected that the anisotropy of such systems may get modified. It has been reported that the presence of a C$_{60}$ layer on Co ultrathin film can control the anisotropy symmetry of the system through an inverse spin reorientation transition from in-plane to out-of-plane via the local hybridization between C$_{60}$ $p_Z$ and Co $d_{Z^{2}}$ orbitals \cite{Bairagi2015,Bairagi2018}. However, the effect of hybridization on the global anisotropy symmetry for in-plane magnetized thin films has not been explored so far.
\\ In this paper, the magnetization reversal has been studied in terms of domain images for a system having Co/C$_{60}$ spinterface and compared to that of its parent single layer Co thin film. The induced moment in the interfacial C$_{60}$ layer is quantified by polarized neutron reflectivity measurement. Further, the effect of such magnetically coupled interface on the net anisotropy of the system has been studied using ferromagnetic resonance technique. 

Co and C$_{60}$ layers have been prepared using dc magnetron sputtering and thermal evaporation techniques, respectively, in a multi-deposition high vacuum chamber manufactured by Mantis Deposition Ltd., UK. The base pressure of the system was better than 5 $\times$ 10$^{-8}$ mbar. The deposition pressure for Co and C$_{60}$ layers were 5 $\times$ 10$^{-3}$ and ~1 $\times$ 10$^{-7}$ mbar, respectively. Co and C$_{60}$ layers have been deposited in-situ i.e. without breaking the vacuum in between the deposition of two consecutive layers to avoid the oxidation and surface contamination of the Co layer. The rate of depositions of Co and C$_{60}$ layers were 0.02 and $\sim$ 0.01-0.013 nm/s, respectively. Before depositing the Co layer, 5 nm of Ta has been deposited as a seed layer using dc sputtering on the Si (100) substrate to promote a better growth of Co. The lattice constant of Ta (a$_{Ta}$) is more than that of Co (a$_{Co}$) and less than Si (a$_{Si}$). As a consequence, the growth of Co is better on Ta than depositing directly on the Si (100) substrate. To prevent oxidation of Co and surface corrosion of C$_{60}$, a Ta capping layer of 3 nm has been deposited in-situ for both the samples. The sample structures are as follows:
\\ Sample 1: Si (100)/Ta(5 nm)/Co(10 nm)/Ta(3 nm)
\\ Sample 2: Si (100)/Ta(5 nm)/Co(10 nm)/C$_{60}$(12 nm)/Ta(3 nm)
\\ C$_{60}$ has been deposited normal to the substrate. However, the Co target is at 30$^{\circ}$ angle w.r.t. the substrate normal due to the in-built geometry of our deposition system. Therefore, a uniaxial anisotropy is expected in our samples due to oblique angle of deposition \cite{Mallik2014,Mallik2018,Mallik2017,Mallik2018_2,Mallick2018}.  The hysteresis loops and the corresponding domain images have been measured at room temperature using magneto optic Kerr effect [MOKE] based microscopy manufactured by Evico magnetics GmbH, Germany \cite{Evico}. The hysteresis measurements have been performed within a field range of 20 mT by varying the angle $\phi$ in longitudinal mode. Here, $\phi$ denotes the angle between the easy axis and applied field direction. Polarized neutron reflectivity [PNR] has been performed at room temperature on the bilayer film (sample 2) along the easy axis at MARIA reflectometer at FRM II, Garching, Germany \cite{Mattauch2018}. The wavelength ($\lambda$) of the neutrons during the PNR measurements has been chosen to be 0.65 nm. The momentum transfer ($Q_Z$) has been measured by rotating the sample in a specific angle range and keeping the wavelength of the neutron ($\lambda$) constant. The non-spin flip [NSF] scattering cross sections  ($R^{++}$) and  ($R^{--}$) have been measured where the first and second signs in the scattering cross section correspond to the polarization of the incident and the reflected neutrons, respectively. A small guiding field has been applied to maintain the polarization of the incident neutrons at MARIA. To evaluate the magnetic moment in the single layer sample 1 and to compare the change in magnetization due to the presence of the spinterface, hysteresis loop has been measured at room temperature within 250 mT external field using superconducting quantum interference device [SQUID] magnetometer manufactured by Quantum Design, USA \cite{SQUID}. To quantify the anisotropy constants and to observe the effect of the Co/C$_{60}$ interface on the anisotropy symmetry, ferromagnetic resonance [FMR] measurements have been performed using Phase FMR spectrometer manufactured by NanOsc AB, Sweden \cite{FMR}. The angle dependent FMR measurements have been performed at a fixed frequency of 10 GHz on both the samples by varying $\phi$ at an interval of 10$^{\circ}$.

\begin{figure}[h!]
	\centering
	\includegraphics[width=0.9\linewidth]{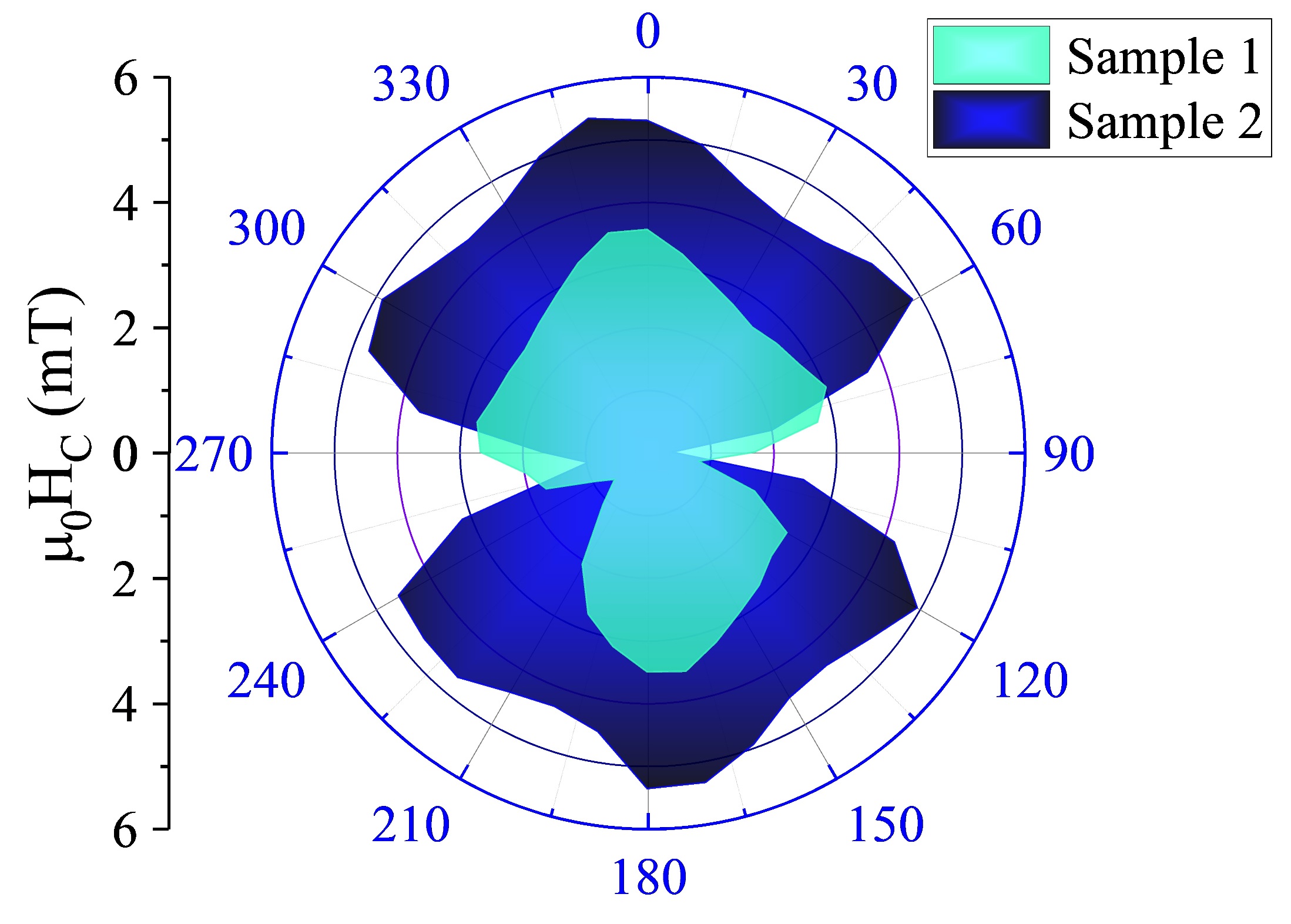}
	\caption{Angle dependent coercivity ($H_C$) plots for sample 1 and 2 to compare the anisotropy symmetry.}
	\label{fig1}
\end{figure}

The angle dependent hysteresis loops for both the samples measured in longitudinal mode at room temperature are shown in figure S1. A qualitative figure to understand the anisotropy symmetry of the samples have been plotted by extracting the angle dependent coercivity ($H_C$) values. The anisotropy symmetry of both the samples is shown in Fig. \ref{fig1}. It should be noted that Co was deposited under oblique angle of incidence in our sputtering chamber. Because of this, the grains of Co are expected to form chain like structure and become elongated towards a specific direction and therefore uniaxial anisotropy is observed in the samples \cite{Mallik2018,Mallick2018,Mallik2014,Mallik2017,Mallik2018_2}. The long range dipolar interaction is the main cause of such elongation of grains and as a result a uniaxial anisotropy is induced in the system \cite{Cherifi2007,Bubendorff2006,Shim2007}. It should be noted that the easy axis of the induced uniaxial anisotropy lies in-plane at 90$^{\circ}$ angle with respect to the projection of the plume direction. For sample 1, the symmetry of the uniaxial anisotropy is not very profound. A probable cause of the asymmetric shape of the anisotropy symmetry seen in Fig. $\ref{fig1}$ can be local dispersion in the anisotropy due to the polycrystalline growth of Co on Ta. Due to the polycrystalline nature of the Co layer, misalignment of local anisotropy in between the neighbouring grains may be present. Hence, a local decrease in net anisotropy can be observed in those regions. 

\begin{figure*}
	\centering
	\includegraphics[width=1.0\linewidth]{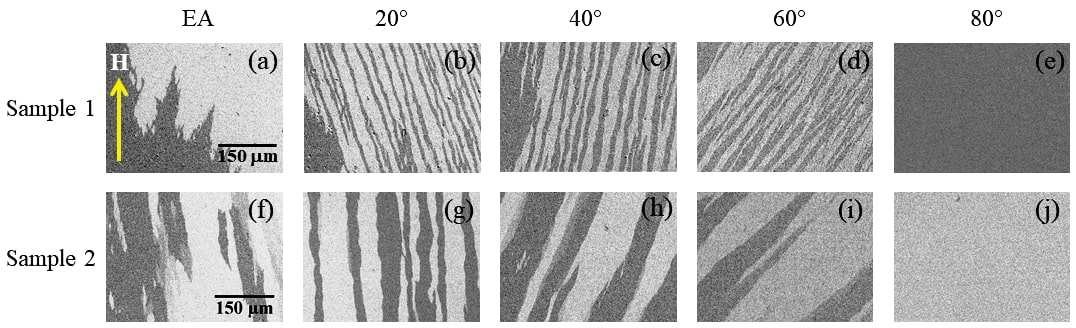}
	\caption{Domain images near the coercivity for samples 1 and 2 are shown in (a) - (e) and (f) - (j), respectively. The domain images are measured using magneto optic Kerr effect [MOKE] based microscopy at room temperature in longitudinal mode by varying the angle ($\phi$) between the easy axis and the applied field direction. The scale bars of the images for samples 1 and 2 are shown in image (a) and (f), respectively. The applied field direction shown in image (a) was kept constant during all the measurements and the sample was rotated accordingly.}
	\label{fig2}
\end{figure*}

It should be noted that $H_C$ increases for sample 2 in comparison to sample 1. This can be explained due to the orbital hybridization at the interface of Co/ C$_{60}$ in sample 2. It is expected from the electronic structures of Co and C$_{60}$ that hybridization is probable between \textit{d} and \textit{p} orbitals of Co and C atoms, respectively. Due to such hybridization, the C$_{60}$ molecules at the interface may exhibit ferromagnetism which may promote the enhancement in $H_C$ in sample 2. Detailed information on the induced magnetism in C$_{60}$ is discussed later in this paper. The presence of local minima and maxima other than 90$^{\circ}$, 270$^{\circ}$ and 0$^{\circ}$, 180$^{\circ}$, respectively, in the anisotropy plot (Fig. $\ref{fig1}$) indicates the increase in dispersion or appearance of another anisotropy in sample 2 in comparison to that of the sample 1. The roughness at the Co/C$_{60}$ interface is expected to be high due to the amorphous growth of C$_{60}$. As a result, the anisotropy becomes more dispersed in sample 2. Further, the variance in $H_C$ values while changing $\phi$ is quite prominent for sample 2 (Fig. $\ref{fig1}$). This indicates that the anisotropy increases in sample 2 in comparison to sample 1 due to the presence of Co/C$_{60}$ interface. It should be noted that the hybridization at the Co/C$_{60}$ interface may promote an interfacial anisotropy which in turn can increase the strength of the uniaxial anisotropy or induce another anisotropy in the bilayer system.

In order to understand the effect of C$_{60}$ on the magnetization reversal of the bilayer sample we have performed domain imaging via Kerr microscopy. The domain images captured near the $H_C$ measured along different angles ($\phi$) for both the samples are shown in Fig. $\ref{fig2}$. The first row corresponds to the domain images taken for sample 1. Along the EA, large branch domains are observed. By rotating $\phi$ away from the EA, domain size decreases and stripe domains are observed. It is known that branch domains are observed in the samples having poor growth. Similarly, stripe domains are characteristics of films having dispersed uniaxial anisotropy due to strain \cite{Hubert1998}. Therefore, due to the polycrystalline growth of Co, branch and stripe domains are observed for different angles. In case of thin films having well defined uniaxial anisotropy, the domain size decreases monotonically from EA towards HA. However, due to the presence of dispersion in anisotropy symmetry, the domain size increases in between $30^{\circ} < \phi < 60^{\circ}$. The variation of average domain widths with respect to $\phi$ is shown in Fig. S2 of the Supplemental Material. It should be noted that the domain widths are calculated by taking the average of the width of the branches/stripes measured at several positions of an image. Therefore, the source of error is the systematic error in the measurement i.e. the least count of the measuring tool. As discussed earlier, the dispersion in anisotropy can be attributed to the fact that the growth of Co is polycrystalline on Ta in our samples.\\
The second row of Fig. $\ref{fig2}$ corresponds to the domain states measured for sample 2 for different angles of rotation ($\phi$). For this sample also, the nature of the domains stipulates dispersion in the anisotropy due to the presence of Co/C$_{60}$ interface. However, the increase in domain size in the bilayer sample indicates the presence of orbital hybridization at the Co/C$_{60}$ interface. For $\phi > 70^{\circ}$, domains are not observed during the reversal as it happens probably via coherent rotation. Therefore, a gradual change in the contrast has been observed for $\phi > 70^{\circ}$ while the field is swept from the positive to negative saturation. Similar to the single layer sample 1, the domain width increases away from EA for the bilayer sample 2. Due to the presence of more dispersion or another anisotropy contribution in sample 2, the domain width at $\phi$ = 40$^{\circ}$ becomes similar to that of at EA. The change in the domain width with respect to $\phi$ for sample 2 is depicted in Fig. S2 of the Supplemental Material. In order to endorse the reproducibility of the results, three sets of samples with similar growth conditions have been measured and verified. 

\begin{figure*}
	\centering
	\includegraphics[width=0.8\linewidth]{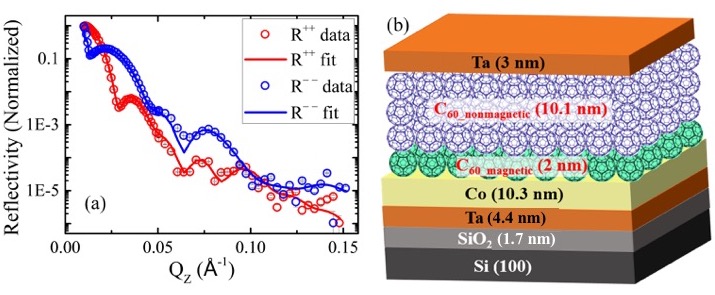}
	\caption{(a) Polarized neutron reflectivity [PNR] data and the corresponding fits for sample 2. The red and blue open circles represent the data measured for the $R^{++}$ and $R^{--}$ channels, respectively. The solid lines correspond to their respective fits. The measurement was performed at the saturation state at room temperature. (b) A schematic representation of the sample structure obtained by fitting the PNR data is shown in (a). The numbers written in brackets in each layer correspond to the fitted thickness of the respective layer.}
	\label{fig3}
\end{figure*}

 PNR has been performed on sample 2 to quantify the induced moment in C$_{60}$ and thickness of the spinterface.  Figure \ref{fig3}(a) shows the PNR data and the corresponding fit measured at positive saturation ($\mu_0H$ = 100 mT) of the sample at room temperature for sample 2. The red and blue open circles represent the data obtained for the $R^{++}$ and $R^{--}$ reflectivities, respectively. The data has been fitted using GenX software \cite{GenX} which is based on Parratt formalism \cite{Parratt1954}. The sample structure obtained from the best fit is shown in Fig. \ref{fig3}(b) with thicknesses extracted for all the layers. The figure of merit of the fit is less than 8E-2. From the fitting, it has been observed that the Co layer exhibits magnetic moment of 1.6 $\pm$ 0.01 $\mu_B$/atom. The magnetic moment of the single layer Co sample 1 has been calculated to be  1.7 $\pm$ 0.03 $\mu_B$/atom from the SQUID loop (Fig. S3 of the Supplemental Material). Therefore, a reduction of $\sim$ 6$\%$ of magnetic moment in Co is observed in the bilayer sample. On top of the Co layer, $\sim$ 2 $\pm$ 0.18 nm of pure C$_{60}$ shows magnetic moment of 0.8 $\pm$ 0.2 $\mu_B$/cage. Further, rest of the 10 $\pm$ 0.11 nm of C$_{60}$ exhibits its inherent diamagnetic behaviour. The induction of moment in C$_{60}$ cage and loss of moment in Co can be explained due to the hybridization between $d_{Co}$ and $p_{C_{60}}$ orbitals. The unpaired electrons from \textit{d} orbital of Co atom can be transferred to the \textit{p} orbital of C atom leading to hybridization. It is highly probable that the density of states of the C$_{60}$ gets strongly modified at the interface resulting in induced ferromagnetism in fullerene \cite{Sanvito2010,Djeghloul2013}. It should be noted that the thickness of the non-magnetic C$_{60}$ layer does not contribute in the magnetic properties of the system \cite{Mallik2019}. The PNR measurement near the coercive field (Fig. S4 of the Supplemental Material) elucidates the anti-parallel coupling between C$_{60}$ and Co layer at the interface. In this case, 17$\%$ of the Co spins are in reversed state and rest are along the field direction. However, 63$\%$ of the magnetic C$_{60}$ spins point in the reverse direction. Therefore, it has been confirmed from the PNR measurement that the change in domain structure and the anisotropy symmetry discussed earlier is a result of the formation of spinterface between Co and C$_{60}$.
 
 \begin{figure}
	\centering
	\includegraphics[width=0.8\linewidth]{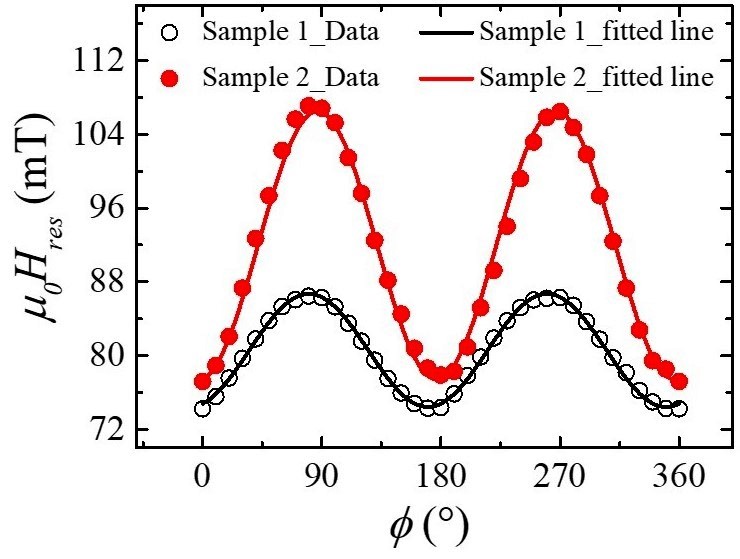}
	\caption{Angle dependent resonance field ($H_{res}$) plot for samples 1 (open circles) and 2 (solid circles) to evaluate the anisotropy constants of the system. The FMR measurements were performed at room temperature by keeping the measurement frequency constant at 10 GHz.}
	\label{fig4}
\end{figure}
 
To quantify and compare the anisotropy in the samples, angle dependent ferromagnetic resonance [FMR] measurement has been performed (Fig. S5 in Supplemental Material). The angle ($\phi$) has been rotated in 10$^{\circ}$ interval for both the samples at a constant frequency of 10 GHz. The corresponding resonance field ($H_{res}$) has been recorded for each angle. Figure \ref{fig4} shows the $H_{res}$ vs $\phi$ data (open black circles for sample 1 and solid red circles for sample 2) and corresponding fits (black solid line for sample 1 and red solid line for sample 2) for both the samples.
The energy equation for a system having uniaxial anisotropy can be written as \cite{Gieniusz2007}:
\begin{equation}
\begin{aligned}
E = & -HM_S[sin\theta sin\theta_M cos(\phi_M - \phi)+cos\theta cos\theta_M] \\
& -2\pi M^2_S sin^2 \theta_M + K_U sin^2 \theta_M + K_1 sin^2 \theta_M sin^2 \phi_M  
\end{aligned}
\end{equation}
Here, $\phi$ is the in-plane angle between the easy axis and the applied field direction. $\phi_M$ is the in-plane angle between the easy axis and the projection of magnetization. $\theta$ and $\theta_M$ are the angles between the z-axis w.r.t. the applied field direction and the magnetization direction, respectively,(see the schematic shown in Fig. S6 in Supplemental Material). However, $\theta_M$ and $\theta$ are considered to be 90$^{\circ}$ as both the magnetization and the field direction lie in the sample plane. Here, the first term of equation 1 corresponds to the total Zeeman energy of the system. Due to the interaction of the external magnetic field ($H$) and the total magnetization of the sample ($M_S$), the in-plane and out-of-plane components of Zeeman energy are $Hsin\theta M_S sin\theta_M cos(\phi_M-\phi)$ and $Hcos\theta M_S cos\theta_M$, respectively. Further, The demagnetization field is proportional to the magnetic free pole density and can be expressed as 2$\pi M_S^2 sin^2 θ\theta_M$. Last two terms of equation 1 correspond to the uniaxial magnetocrystalline anisotropy energy where $K_1$ is the in-plane uniaxial anisotropy constant and $K_U$ is the perpendicular contribution of the anisotropy.

The dispersion relation i.e. the fitting equation to evaluate the strength of the anisotropies present in both the samples can be derived from equation 1 and can be written as \cite{Gieniusz2007}:
\begin{equation}
\begin{aligned}
(\omega / \gamma)^2 = &[Hcos(\phi_M - \phi) - h_U + h_1 sin^2 \phi_M]\\&[Hcos(\phi_M - \phi) 
- h_1 + 2h_1 sin^2 \phi_M]
\end{aligned}                      
\end{equation}      
where, $h_1$ is the in-plane anisotropy field and can be expressed as $h_1 = 2(K_1 / M_S)$. By fitting the angle dependent $H_{res}$ values (Fig. \ref{fig4}) with equation 2, the anisotropy constant $K_1$ is extracted to be $8.78 \times 10^3$ and $1.71 \times 10^4 J/m^3$ for samples 1 and 2, respectively. Therefore, it is concluded that the anisotropy increases by two times in the Co/C$_{60}$ bilayer sample 2 than that of the single layer Co sample 1. Hence, the orbital hybridization at the interface of Co and C$_{60}$  may lead to deformation/elongation of the shape of orbitals which in turn promotes the enhancement of global anisotropy in bilayer system by probably inducing an extra interfacial anisotropy. The origin and nature of this interfacial anisotropy needs to be elucidated by future density functional theory calculations.

In summary, the magnetic properties of bilayer Co/C$_{60}$ sample have been studied and compared with its parent single layer Co thin film in this paper. Formation of $\sim$2 $\pm$ 0.18 nm spinterface is observed at the Co/C$_{60}$ interface which exhibits 0.8 $\pm$ 0.2 $\mu_B$/cage magnetic moment. The domain size as well as the coercivity of the bilayer sample increases due to the presence of the spinterface. The anisotropy is enhanced by two times in the bilayer sample in comparison to its single layer counterpart due to the $p-d$ orbital hybridization between the C and Co atoms. The possibility of tuning the magnetic anisotropy might have a profound impact in improving the versatility of devices like magnetic tunnel junctions.

\section*{SUPPLEMENTARY MATERIAL}
 {See supplementary material for the angle dependent Kerr loops (fig. S1), domain width comparison (fig. S2), M-H loops from SQUID measurement (fig. S3), PNR data near the coercive field (fig. S4), angle dependent FMR spectra (fig. S5) and schematic for geometry of the angle dependent FMR measurement (fig. S6).}

The authors thank Department of Atomic Energy, and Department of Science and Technology - Science and Engineering Research Board (SB/S2/CMP-107/2013 and EMR/2016/007725), Govt. of India, for providing the financial support to carry out the experiments. The authors also thank the Department of Science and Technology, India (SR/NM/Z-07/2015) for the financial support for performing the neutron experiments and Jawaharlal Nehru Centre for Advanced Scientific Research [JNCASR] for managing the project.

\end{document}